\documentclass[nofootinbib,showpacs,prd,superscriptaddress]{revtex4}

\usepackage{amsmath}
\usepackage{amsfonts}
\usepackage{amssymb}
\usepackage{graphicx}

\begin{document}


\title{Cardy-Verlinde entropy in Ho\v rava-Lifshitz gravity}

\author{Orlando Luongo}
\email{luongo@na.infn.it}
\affiliation{Department of Mathematics and Applied Mathematics, University of Cape Town, South Africa, Rondebosch 7701, Cape Town, South Africa.}
\affiliation{Astrophysics, Cosmology and Gravity Centre (ACGC), University of Cape Town, Rondebosch 7701, Cape Town, South Africa.}
\affiliation{Dipartimento di Fisica, Universit\`a di Napoli ''Federico II'', Via Cinthia, I-80126, Napoli, Italy.}
\affiliation{Istituto Nazionale di Fisica Nucleare (INFN), Sez. di Napoli, Via Cinthia, Napoli, Italy.}

\author{Giovanni Battista Pisani}
\affiliation{Dipartimento  di Fisica, ``Sapienza" Universit\`a di Roma, Piazzale Aldo Moro 5, I-00185, Roma, Italy.}

\author{Hernando Quevedo}
\email{quevedo@nucleares.unam.mx}
\affiliation{Instituto de Ciencias Nucleares, Universidad Nacional Aut\'onoma de M\'exico, Mexico.}
\affiliation{Dipartimento  di Fisica, ``Sapienza" Universit\`a di Roma, Piazzale Aldo Moro 5, I-00185, Roma, Italy.}

\begin{abstract}
We investigate homogeneous cosmological models with perfect-fluid sources in the framework of the Ho\v rava-Lifshitz model for quantum gravity.
We show that the Hamiltonian constraint of such spacetimes can be rewritten as the Cardy formula for the entropy in conformal field theory.
The Cardy entropy is shown to depend explicitly on the value of the Ho\v rava parameter $\lambda$ so that it can be interpreted as determining
the entropy and the gravitational interaction of the theory. Moreover, we show that Verlinde's Pythagorean representation of the Hamiltonian constraint is
also valid in the case of homogeneous Ho\v rava-Lifshitz spacetimes. We interpret these results as a further indication of a deep relationship between gravity, thermodynamics and holography in the quantum regime.
\end{abstract}


\maketitle

\section{Introduction}

All the efforts spent to develop a self-consistent theory of quantum gravity have been so far incomplete \cite{prima}. Even though many approaches have been carried forward during the last few decades, the problem of obtaining a covariant scheme for quantum gravity still remains open and, consequently, many phenomenological proposals have been discussed in the literature \cite{secondo,terzo,quarto,quinto}.
An appealing technique consists in assuming the violation of the Lorentz invariance at some particular energy regimes \cite{sesto}. The idea is that Lorentz invariance is somehow broken at ultraviolet  scales, providing quantum effects which have consequences on the observable universe itself.
Recently, following this philosophy Ho$\check{\textrm{r}}$ava proposed an alternative theory which reduces to Einstein's gravity with a cosmological constant in the infrared  regime, leading to complicated  modifications at the ultraviolet regime \cite{horava}. The underlying philosophy of Ho\v rava's model is to include Lorentz breaking terms in the action which provides a different scaling between space and time at a ultraviolet fixed point.
The transformation  $x^{i}\rightarrow lx^{i},\quad t\rightarrow l^{z}t$, where $z$ is the scaling exponent, in the case $z=3$, leads to a theory which is renormalizable  by power counting. This idea was first proposed by Lifshitz \cite{teoriadilif} and, in the case of the Ho\v rava approach, it leads to a
model which is non-relativistic in the ultraviolet regime.
However, in the infrared limit, the model manifests an emerging 4-dimensional general covariance.
Due to these considerations, the Ho$\check{\textrm{r}}$ava framework is frequently referred to as the {Ho$\check{\textrm{r}}$ava-Lifshitz (HL) model}.
Although the model is clearly appealing to describe quantum effects in gravity, its physical interpretation is still open.
For instance, the model depends upon several parameters, each of them having a precise physical role. Nevertheless, the  interpretation of those parameters is not completely known. In this work, we will focus on the parameter $\lambda$ which is usually fixed in the infrared limit in order to obtain Einstein's gravity.

In this work, we will consider homogeneous spacetimes with a perfect fluid as the gravitational source under the condition that they satisfy the field equations of the Ho\v rava-Lifshitz model. In particular, we will explicitly analyze the isotropic Friedmann-Lema{\^i}tre-Robertson-Walker (FLRW) spacetime and the anisotropic Bianchi cosmologies of the type I, V and IX, following the procedure formulated in \cite{fq07}.
 We first derive the field equations and write them in a form which is suitable for our analysis. In particular, we express the Hamiltonian constraint in terms of the Hubble parameters which determine the expansion in different spatial directions.
Then, we assume the validity of the Cardy entropy which was originally derived in 2-dimensional conformal field theory. Assuming the validity of the Cardy formula implies that we are assuming a holographic treatment of the entropy in the HL model. If this assumption turns out to be true in the corresponding cosmological scenarios, we can conclude that the holographic principle can be used in the corresponding models. In fact, we will show that for all homogeneous spacetimes under consideration in this work it is possible to define the central charge and the eigenvalue of the Virasoro operator in such a way that the Cardy formula is identically satisfied, generating an explicit expression for the Cardy entropy. In doing so, it turns out that the Ho\v rava parameter $\lambda$ enters explicitly the Cardy entropy, allowing us to interpret is determining the entropy and the gravitational interaction of the theory. Moreover, we will calculate the Hubble, Bekenstein and Bekenstein-Hawking for the anisotropic Bianchi IX cosmological model, and will show that they can be arranged in such a way that the corresponding Hamiltonian constraint can be cast into the Pythagorean representation proposed recently by Verlinde \cite{ver}.

This paper is structured as follows. In Sec. \ref{sec:hor}, we present a brief review of the Ho\v rava action and comment on the free parameters that enter the theory.
In Sec. \ref{sec:cosm}, we study homogeneous spacetimes with a perfec-fluid source. We consider the FLRW metric and the Bianchi I, V and IX metrics for which we derive the field equations explicitly. Special attention is given to the form of the Hamiltonian constraint which is expressed in a compact form in terms of the scale factors, the Hubble parameters and the parameters entering the HL model. In Sec. \ref{sec:cve}, we analyze the Hamiltonian constraints of all the homogeneous cosmological spacetimes and establish a unique relationship that allows us to define the Cardy entropy for each spacetime. We also show that the Pythagorean representation of the constraints in terms of entropies is valid in the HL model. Finally, in Sec. \ref{sec:con}, we discuss our results.

\section{The Ho\v rava-Lifshitz paradigm}
\label{sec:hor}

Consider an arbitrary 4-dimensional metric in the framework of the ADM (Arnowitt-Deser-Misner) formalism, i.e.,
\begin{equation}
ds^{2}=-N^{2}c^{2}dt^{2}+g_{ij}(dx^{i}-N^{i}dt)(dx^{j}-N^{j}dt)\,,
\end{equation}
where we introduced the lapse function $N$, the shift vector $N^{i}$ and the
3-dimensional metric $g_{ij}$.  The Einstein-Hilbert covariant action in the ADM-decomposition is given by
\begin{equation}
S_{EH}=\frac{1}{16\pi G}\int
d^{4}x\sqrt{g}N\left(K_{ij}K^{ij}-K^{2}+R-2\Lambda\right)\,,
\end{equation}
where
\begin{equation}
K_{ij}=\frac{1}{2N}(\dot{g}_{ij}-\nabla_{i}N_{j}-\nabla_{j}N_{i})\,
\end{equation}
is the extrinsic curvature tensor and $K$ its trace, where the dot represents the time derivative.

The HL quantum gravity model is based upon a generalization of the Einstein-Hilbert action which breaks covariance in order to obtain in the ultraviolet limit
a theory that is renormalizable by power counting. Then, the Ho\v rava action can be expressed as \cite{horava}
\begin{equation}\label{causaimpossibile}
S=\int
dtd^{3}x(\mathcal{L}_{0}+\mathcal{L}_{1})\,,
\end{equation}
with
\begin{equation}
\mathcal{L}_{0}=\sqrt{g}N\left\{
\frac{2}{\kappa^{2}}\left(K_{ij}K^{ij}-\lambda
K^{2}\right)+\frac{\kappa^{2}\mu^{2}(\Lambda_{W}R^{(3)}-3\Lambda_{W}^{2})}{8(1-3\lambda)}\right\}\,,
\end{equation}
and
\begin{equation}
\mathcal{L}_{1}=\sqrt{g}N\left\{
\frac{\kappa^{2}\mu^{2}(1-4\lambda)}{32(1-3\lambda)}(R^{(3)})^{2}-\frac{\kappa^{2}}{2\nu^{4}}\left(C_{ij}-\frac{\mu\nu^{2}}{2}R_{ij}^{(3)}\right)\left(C^{ij}-\frac{\mu\nu^{2}}{2}R^{(3)ij}\right)\right\}\,,
\end{equation}
where  $R_{ij}^{(3)}$ and $R^{(3)}$ are the Ricci tensor and the scalar curvature for the 3-dimensional metric $g_{ij}$, respectively, and
\begin{equation}
C^{ij}=\epsilon^{ikl}\nabla_{k}\left(R_{\quad
l}^{(3)j}-\frac{1}{4}R^{(3)}\delta_{l}^{j}\right)\,,
\end{equation}
is known in literature as the Cotton tensor. The free parameters entering the Ho\v rava action  ($\ref{causaimpossibile}$) are: $\lambda$, $\kappa$, $\mu$,
$\nu$ and $\Lambda_{W}$. Soon after the publication of the model, it was found that the
Schwarzschild-AdS black hole solution is not recovered in the infrared
limit, although Einstein's theory with cosmological model was
obtained at the level of the action \cite{BlackHoles}. This
difficulty was solved by introducing an additional parameter which
modifies the IR behavior \cite{GeneralizzazioneHorawa,adddreee}. In this work, however, we will use the original Ho\v rava action since our results do not depend on
the value of the additional parameter.

The five constants $\kappa$, $\lambda$, $\mu$,
$\nu$, and $\Lambda_W$ are parameters which determine the
velocity of the light $c$, the gravitational constant $G$ and the
Einstein cosmological constant $\Lambda$ by means of \cite{Muinpark}
\begin{equation}
\label{uno}
c^2 = \frac{\kappa^4\mu^2 |\Lambda_W|}{8(3\lambda-1)^2}, \quad
G=\frac{\kappa^2c^2}{16\pi(3\lambda-1)}\quad
\Lambda_W=\frac{2}{3}\Lambda.
\end{equation}
These three constraints imply that three constants of the HL model can be fixed by using experimental data.
Notice that for $\lambda < 1/3$ the gravitational constant becomes
negative, indicating the presence of repulsive gravity. We therefore
limit ourselves to the case $\lambda
> 1/3$. Consequently, the HL model possesses only two free
parameters which should be chosen in accordance with observations. In this work, we will focus on the investigation of the physical significance of the
parameter $\lambda$ in the context of cosmological models.

\section{Cosmological dynamics}
\label{sec:cosm}

The simplest cosmological models assume a high degree of symmetry in order to be able to handle the corresponding field equations. In this section, we will assume
that the spacetime is homogeneous so that we are left with only two possible classes of models, namely, isotropic and anisotropic cosmological models. In the first case,   the geometry of the spacetime is described by the FLRW metric
\begin{equation}
ds^{2}=-dt^{2}+a(t)^{2}\left[\frac{dr^{2}}{1-kr^{2}}+r^{2}\left(d\theta^{2}+\sin^{2}\theta
d\phi^{2}\right)\right]\ ,
\label{flrw}
\end{equation}
where $a(t)$ is the scale factor and $k=-1,0,1$. The corresponding field equations can be derived from the variation of the Ho\v rava action. We obtain
\begin{equation}\label{ocas}
H^{2}=\frac{2}{3\lambda-1}\left[\frac{8\pi G}{3}\rho_{tot}-\frac{k}{a^{2}}-\frac{2k^{2}}{3\lambda-1}\left(\frac{4\pi G\mu}{a^{2}}\right)^{2}\right]\,,
\end{equation}
\begin{equation}
\frac{\ddot{a}}{a}=\frac{2}{3\lambda-1}\left[-\frac{4\pi G}{3}(\rho_{tot}+3p_{tot})+\frac{2k^{2}}{3\lambda-1}\left(\frac{4\pi G\mu}{a^{2}}\right)^{2}\right]\,,
\end{equation}
where
the total density and pressure  are defined as $\rho_{tot}=\rho_{\Lambda_{W}}+\rho$ and $p_{tot}=p_{\Lambda_{W}}+p$, respectively, with
$\rho_{\Lambda_{W}}\equiv\frac{3\Lambda_{W}}{16\pi G}$ and
$p_{\Lambda_{W}}\equiv-\rho_{\Lambda_{W}}$.
The limiting case of the
$\Lambda$CDM model is recovered when $\lambda\rightarrow1$ and $\mu\rightarrow0$.

For later use, it is convenient to rewrite the Hamiltonian constraint ($\ref{ocas}$)  as
\begin{equation}
\frac{3\lambda-1}{2}H^{2}+\frac{k+M}{a^{2}}=\frac{8\pi G}{3}\rho_{tot},
\label{hamflrw}
\end{equation}
where $M=M(a, \lambda, \mu)$ is a function of the scale factor and the Ho$\check{\textrm{r}}$ava parameters. The explicit value of this function depends on the topology of the cosmological model and can be expressed as given in Table \ref{table1}.

\begin{table}
\begin{center}
\begin{tabular}{|c|c|c|c|}
\hline FLRW universe & $ \quad k \quad$ & $\qquad M \qquad$
\tabularnewline
\hline
flat & $0$ & $0$
\tabularnewline
open & $-1$ & $\frac{2}{3\lambda-1}\left(\frac{4\pi G\mu}{a}\right)^{2}$
\tabularnewline
close & $+1$ & $\frac{2}{3\lambda-1}\left(\frac{4\pi G\mu}{a}\right)^{2}$
\tabularnewline
\hline
\end{tabular}
\caption{Hamiltonian constraint (\ref{hamflrw}) for isotropic spacetimes}
\label{table1}
\end{center}
\end{table}

We now consider the case of anisotropic cosmological models. We will restrict ourselves to the study of
the Bianchi I, V and IX models which are described by the metrics
\begin{equation}
\label{metricI}
^{(I)}ds^{2}=-dt^{2}+a_{1}^{2}dx^{2}+a_{2}^{2}dy^{2}+a_{3}^{2}dz^{2}\,,
\end{equation}
\begin{equation}
\label{metricV}
^{(V)}ds^{2}=-dt^{2}+a_{1}^{2}dx^{2}+e^{2x}a_{2}^{2}dy^{2}+e^{2x}a_{3}^{2}dz^{2}\,,
\end{equation}
\begin{equation}
\label{metricIX}
^{(IX)}ds^{2}=-dt^{2}+a_{1}^{2}(\omega_{1})^{2}+a_{2}^{2}(\omega_{2})^{2}+a_{3}^{2}(\omega_{3})^{2}\,,
\end{equation}
where the scale factors $a_{i}$ ($i=1,2,3$) depend on time only, and
\begin{equation}
 \begin{array}{l}
\omega_{1}=\frac{1}{2}\left(-dx\sin z+dy\sin x\cos z\right)\,,\\
\omega_{2}=\frac{1}{2}\left(dx\cos z+dy\sin x\sin z\right)\,,\\
\omega_{3}=\frac{1}{2}\left(dy\cos x+dz\right)\,.
\end{array}
\end{equation}

In the case of the Bianchi IX metric, it is convenient to introduce the alternative representation
\begin{equation}
\label{metricIXmodif}
^{(IX)}ds^{2}=-dt^{2}+e^{-2\Omega}\left[e^{2X+2Y}(\omega_{1})^{2}+e^{2X-2Y}(\omega_{2})^{2}+e^{-4X}(\omega_{3})^{2}\right]\,.
\end{equation}
which is useful for concrete calculations in terms of the scale factors
\begin{equation}
a_{1} = e^{-\Omega+X+Y}\,,\quad a_{2} = e^{-\Omega+X-Y}\,,\quad a_{3} = e^{-\Omega-2X}\,.
\end{equation}

Introducing the directional Hubble parameters $H_{i}=\dot{a_{i}}/a_{i}$, the computation of the corresponding  Hamiltonian constraint yields
\begin{equation}
\frac{1}{6}(\lambda-1)(H_{1}^{2}+H_{2}^{2}+H_{3}^{2}) +\frac{\lambda}{3}(H_{1}H_{2}+H_{1}H_{3}+H_{2}H_{3})+\frac{k+F}{a_{1}^{2}}=\frac{8\pi G}{3}\rho_{tot}\,,
\label{hambi}
\end{equation}
where $F=F(a_{1}, a_{2}, a_{3}, \lambda, \mu)$ depends on the scale factors and the Ho$\check{\textrm{r}}$ava parameters, and its explicit form is given in Table
\ref{table2} where

\begin{table}
\begin{center}
\begin{tabular}{|c|c|c|}
\hline Bianchi type & $k$ & $F$
\tabularnewline
\hline
& &
\tabularnewline
I & $0$ & $0$
\tabularnewline
& &
\tabularnewline
V & $-1$ & $\frac{2}{3\lambda-1}\left(\frac{4\pi G\mu}{a_{1}}\right)^{2}$
\tabularnewline
& &
\tabularnewline
IX &
$\quad +1\quad$ &
$\quad \frac{\epsilon^{2}}{3} +\frac{2}{3\lambda-1}\left(\frac{4\pi G\mu}{a_{1}}\right)^{2}\left[3(3-8\lambda)\left(1+\frac{\epsilon^{2}}{3}\right)^{2}-8(1-3\lambda)\left(1+\frac{\theta^{4}}{3}\right)\right] \quad$
\tabularnewline
& &
\tabularnewline
\hline
\end{tabular}
\caption{Hamiltonian constraint (\ref{hambi}) for anisotropic cosmological models}
\label{table2}
\end{center}
\end{table}
\begin{equation}
\label{epsilonthetadef}
\begin{array}{c}
\epsilon^{2}\equiv1-\frac{a_{3}^{2}}{a_{2}^{2}}-2\left(1-\frac{a_{1}^{2}}{a_{2}^{2}}\right)-\frac{a_{2}^{2}}{a_{3}^{2}}\left(1-\frac{a_{1}^{2}}{a_{2}^{2}}\right)^{2},\\
\theta^{4}\equiv1-\frac{a_{3}^{4}}{a_{2}^{4}}-2\left(1-\frac{a_{1}^{4}}{a_{2}^{4}}\right)-\frac{a_{2}^{4}}{a_{3}^{4}}\left(1-\frac{a_{1}^{4}}{a_{2}^{4}}\right)^{2}.\end{array}
\end{equation}

As for the dynamical equations, we now have three second-order differential equations which relate the second time-derivatives of the scale factors. Although these equations are not strictly necessary for the following analysis, for the sake of completeness and for future use we present them explicitly in the Appendix.


\section{The Cardy-Verlinde entropy}
\label{sec:cve}

There are several ways to treat a cosmological model as a thermodynamic system to which a particular value of entropy can be associated. One can, for instance, demand
that the cosmological model satisfy the first law or thermodynamics $dE = TdS - pdV$ which in the case of homogeneous cosmological models implies that \cite{fq07}
\begin{equation}
T\dot S = V\left[ \dot\rho + (p+\rho) \frac{\dot V}{V}\right] \ ,
\label{firstlaw}
\end{equation}
where we assumed that $E=\rho V$. Then, using the conservation law, an equation of state and Euler's identity, it is possible to obtain explicit expressions for the entropy and the temperature in terms of the volume $V$ or, equivalently, the cosmic time $t$. Although this procedure gives reasonable results in the case of homogeneous Bianchi models \cite{opq03,fq07}, under the presence of inhomogeneities the physical predictions are incompatible with observations \cite{qs,kqs}.

An alternative approach consists in using the holographic principle and the results of conformal field theory (CFT) \cite{suss}.
Indeed, in 2-dimensional CFT it is possible to count the microscopic states of a physical system in a simple manner by applying the Cardy formula \cite{Cardy}
\begin{equation}
S_C = 2 \pi \sqrt{\frac{c}{6} \left(L_0 - \frac{c}{24}\right)}\ ,
\label{augha}
\end{equation}
where $c$ is the central charge and $L_0$ is the eigenvalue of the Virasoro operator. Recently, Verlinde proposed the universal validity of Cardy's entropy and found
an interesting link with Friedmann's equations in general relativity \cite{ver}. Indeed, the Hamiltonian constraint in close FLRW spacetimes turns out to coincide with the formula for the Cardy entropy if the relationships
\begin{equation}
 \begin{array}{c}
S_{C}=\frac{VH}{2G}\,,\\
L_{0}=\frac{aE}{3}\,,\\
c=\frac{3V}{a\pi G}\,,\end{array}
\end{equation}
are satisfied with $V=a^3$.
This formal analogy was confirmed also in the case of more general FLRW spacetimes \cite{youm99} and of Bianchi models \cite{opq03,fq07},
for which the generalized Cardy formula reads
\begin{equation}
S_{C}=2\pi\sqrt{\frac{c}{6}\left(L_{0}-k\frac{c}{24}\right)}\,.
\label{cardyk}
\end{equation}
The validity of this formula in so many different cosmological scenarios can be interpreted as indicating the existence of a deeper relationship between general relativity, thermodynamics and holography.

From the results presented in the previous section, one can easily see that in the  HL quantum gravity model, the Hamiltonian constraints of all FLRW cosmological models satisfy the generalized Cardy formula if the corresponding entropy is identified as
\begin{equation}
S_{C}=\frac{VH}{2G}\sqrt{\frac{3\lambda-1}{2}}\,,
\label{cardy2}
\end{equation}
while the central charge and the eigenvalue of the Virasoro operator must be chosen as given in Table \ref{table3}.  In the limiting case  $\lambda\rightarrow1$ and
$\mu\rightarrow0$, our results reduce to those of Einstein gravity.

\begin{table}
\begin{center}
\begin{tabular}{|c|c|c|c|c|}
\hline FLRW universe & $\qquad k \qquad$ & $S_{C}$ & $\qquad L_{0} \qquad$ & $\qquad c \qquad$
\tabularnewline
\hline
& & & &
\tabularnewline
flat & $0$ & $2\pi\sqrt{\frac{c}{6}L_{0}}$ & $\frac{aE}{3}$ & $\frac{3V}{a\pi G}$
\tabularnewline
& & & &
\tabularnewline
open & $-1$ & $2\pi\sqrt{\frac{c}{6}\left(L_{0}+\frac{c}{24}\right)}$ & $\frac{aE}{3\sqrt{-1-M}}$ & $\frac{3V\sqrt{1-M}}{a\pi G}$
\tabularnewline
& & & &
\tabularnewline
close & +$1$ & $2\pi\sqrt{\frac{c}{6}\left(L_{0}-\frac{c}{24}\right)}$ & $\frac{aE}{3\sqrt{1+M}}$ & $\frac{3V\sqrt{1+M}}{a\pi G}$
\tabularnewline
& & & &
\tabularnewline
\hline
\end{tabular}
\caption{Cardy entropy for FLRW spacetimes in HL gravity}
\label{table3}
\end{center}
\end{table}

We see that the parameter $\lambda$ enters explicitly the expression for the Cardy entropy (\ref{cardy2}), allowing us to investigate its
physical significance. Indeed, the constant factor $\sqrt{(3\lambda -1)/2}$ represents the only difference between Einstein classical gravity
and the HL model for quantum gravity. Thus, the parameter $\lambda$ can be considered as a measure of the classical and quantum gravitational interaction.
For the limiting value $\lambda=1/3$, the Cardy entropy vanishes indicating that no microscopic states can be associated to the corresponding physical system. This is in accordance with the physical interpretation of the HL model since this case corresponds to a vanishing value for the gravitational constant, i.e. no gravitational interaction is present. In fact, the value $\lambda=1/3$ corresponds to the transition point between repulsive gravity ($\lambda<1/3$) and
attractive gravity $(\lambda>1/3$).  For $\lambda<1/3$, the Cardy entropy becomes imaginary which can be interpreted as a manifestation of the absence of physical  microscopic states in a system dominated by repulsive gravity. We interpret this result as an indication that repulsive gravity is not a physical meaningful theory within the HL quantum gravity model. Starting at $\lambda = 1/3$, the attractive gravitational interaction increases as $\lambda$ increases, and for $\lambda=1$ it reaches the classical
value of general relativity. As $\lambda$ increases, the Cardy entropy increases as $\lambda^{1/2}$ and at some point the theory will reach a quantum regime
whose gravitational interaction will depend on the value of $\lambda$. In particular, $\lambda$ is related to measurements of Lorentz invariance deviations and can be constrained at different cosmological regimes to get limits over quantum gravity effects \cite{sotrep}.

We now consider Bianchi cosmologies in HL gravity. As before, the idea is to identify $S_C$, $L_0$ and $c$ such that the Cardy formula (\ref{cardyk}) coincides with the Hamiltonian constraint (\ref{hambi}). One can show that there is a unique identification which satisfies this condition, namely,
\begin{equation}
S_{C}=\frac{V}{2G}\sqrt{\frac{1}{6}(\lambda-1)(H_{1}^{2}+H_{2}^{2}+H_{3}^{2})+\frac{\lambda}{3}(H_{1}H_{2}+H_{1}H_{3}+H_{2}H_{3})}\,,
\end{equation}
while the values of $L_0$ and $c$ depend on the model as given in Table \ref{table4}. In the limiting isotropic case $(H_1=H_2=H_3)$, we obtain the values for the FLRW spacetime given above. This proves that the Cardy formula is also valid for homogeneous spacetimes in HL gravity.

\begin{table}
\begin{center}
\begin{tabular}{|c|c|c|c|c|}
\hline Bianchi type & $\qquad k \qquad$ & $S_{C}$ & $\qquad L_{0} \qquad$ & $\qquad c \qquad$
\tabularnewline
\hline
& & & &
\tabularnewline
I & $0$ & $2\pi\sqrt{\frac{c}{6}L_{0}}$ & $\frac{a_{1}E}{3}$ & $\frac{3V}{a_{1}\pi G}$
\tabularnewline
& & & &
\tabularnewline
V & $-1$ & $2\pi\sqrt{\frac{c}{6}\left(L_{0}+\frac{c}{24}\right)}$ & $\frac{a_{1}E}{3\sqrt{-F-G}}$ & $\frac{3V\sqrt{1-F}}{a_{1}\pi G}$
\tabularnewline
& & & &
\tabularnewline
IX & +$1$ & $2\pi\sqrt{\frac{c}{6}\left(L_{0}-\frac{c}{24}\right)}$ & $\frac{a_{1}E}{3\sqrt{F+G}}$ & $\frac{3V\sqrt{1+F}}{a_{1}\pi G}$
\tabularnewline
& & & &
\tabularnewline
\hline
\end{tabular}
\caption{Cardy entropy for Bianchi cosmologies in HL gravity}
\label{table4}
\end{center}
\end{table}

\begin{figure}
\includegraphics[scale=0.4]{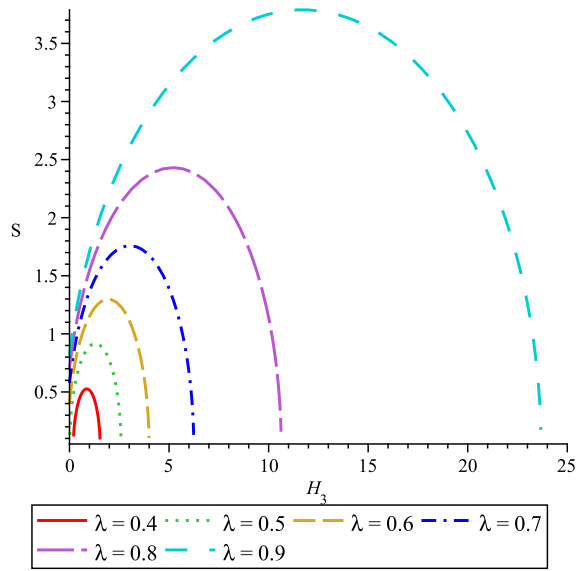}%
\caption{A normalized Cardy entropy as a function of the Hubble parameter $H_3$ for different values of the Ho\v rava parameter $\lambda$.
For concreteness we fix the remaining Hubble parameters as $H_1=0.7$ and $H_2=0.6$.}
\label{fig1}%
\end{figure}

Notice that the validity of the Cardy entropy imposes certain conditions on the Hubble parameters. In fact, for values of $\lambda$ within the
interval $[1,\infty)$ the Hubble parameters $H_1$, $H_2$ and $H_3$ can take any positive real values. This means that the velocity of expansion
can be arbitrarily large in any spatial direction. In the interval $\lambda\in (1/3,1)$, however, the reality condition of the Cardy entropy
implies that the expansion velocity is limited by the value of $\lambda$. This behavior is illustrated in Fig. \ref{fig1}, where we plot the (normalized) Cardy entropy in terms of $H_3$ for different values of $\lambda$ and fixed values of $H_1$ and $H_2$. We see that in order for the Cardy entropy to be
a real quantity, the value of the Hubble parameter $H_3$ must be inside a finite interval whose maximum depends on the value of $\lambda$. This result implies that the expansion velocity in the spatial directions cannot be arbitrarily large. The parameter $\lambda$ acts as a physical barrier which limits the expansion velocity of  an anisotropic universe.

In the case of FLRW spacetimes in Einstein gravity, Verlinde noticed that the Hamiltonian constraint can be written in a Pythagorean form
\begin{equation}
S_{H}^{2} - 2 S_{BH} S_{B} + S_{BH}^2 = 0\ ,
\label{verpyt}
\end{equation}
where $S_H$, $S_{BH}$ and $S_B$ are the Hubble,  Bekenstein-Hawking, and Bekenstein entropy, respectively. A lengthy but straightforward calculation shows that the Hamiltonian constraints (\ref{hamflrw}) and (\ref{hambi}) in HL gravity  can also be put in the Pythagorean form. Indeed, in the case of the Bianchi IX model, the entropies entering Verlinde's formula (\ref{verpyt}) can be expressed as
\begin{equation}
S_{H}=\frac{V}{2G}\sqrt{\frac{\lambda}{3}\left(H_{1}H_{2}+H_{1}H_{3}+H_{2}H_{3}\right)+\frac{\lambda-1}{6}\left(H_{1}^{2}+H_{2}^{2}+H_{3}^{2}\right)} \ ,
\end{equation}
\begin{equation}
S_{BH}=\frac{V}{2Ga_{1}}\sqrt{1+\frac{\epsilon^{2}}{3}+\left(\frac{2}{3\lambda-1}\right)\left(\frac{4\pi G\mu}{a_{1}}\right)^{2}\left[8\left(3\lambda-1\right)\left(1+\frac{\theta^{4}}{3}\right)-3\left(8\lambda-3\right)\left(1+\frac{\epsilon^{2}}{3}\right)^{2}\right]}\ ,
\end{equation}
\begin{equation}
S_{B}=\frac{2\pi}{3}\frac{E a_{1}}{\sqrt{1+\frac{\epsilon^{2}}{3}+\left(\frac{2}{3\lambda-1}\right)\left(\frac{4\pi G\mu}{a_{1}}\right)^{2}\left[8\left(3\lambda-1\right)\left(1+\frac{\theta^{4}}{3}\right)-3\left(8\lambda-3\right)\left(1+\frac{\epsilon^{2}}{3}\right)^{2}\right]}}\ .
\end{equation}
In the isotropic limit, the results are consistent with those obtained for FLRW spacetimes. This form of the Hamiltonian constraint allows us to  represent the dynamical evolution of the entropies as being inside a circle of radius $S_B$. In the case of the FLRW spacetimes, this radius is constant so that the dynamical evolution of the Hubble and Bekenstein-Hawking entropies are restricted to evolve inside the circle. In the anisotropic case, however, $S_B$ depends explicitly on time so that the evolution of the remaining entropies is no longer restricted to be inside a fixed circle, leading to a more complex pictorial representation which physically implies that the evolution depends on the equation of state of the fluid \cite{opq03}. The radius $S_B$  also depends explicitly on the value of the parameter
$\lambda$. This means that for a given value of $\lambda$ the radius of the circle cannot be arbitrarily small or arbitrarily large as time passes, but it
depends explicitly on the value of $\lambda$. In other words, the Ho\v rava parameter $\lambda$ influences the evolution of the entropies in the Verlinde representation.

\section{Final outlooks and perspectives}
\label{sec:con}

In this work, we investigated the structure of the Hamiltonian constraint for homogeneous cosmological models with a perfect-fluid source in the Ho\v rava-Lifshitz quantum gravity model. We calculated explicitly the field equations for the isotropic FLRW metric and for the anisotropic Bianchi I, V and IX metrics. The metrics were chosen in such a way that in each case the resulting Hamiltonian constraint has a particular compact and simple form, which allows us to perform a comparative analysis. We assume the validity of the Cardy entropy formula which is used in 2-dimensional conformal field theory to carry out a counting of the microscopic states of physical systems.

It was shown that the Hamiltonian constraint of homogeneous cosmological spacetimes can be identified with the Cardy entropy, by choosing in a unique way  the central charge and the eigenvalue of the Virasoro operator which enter the Cardy formula. As a result, we obtain explicit expressions for the Cardy entropy of the FLWR and Bianchi spacetimes. This proves the validity of the Cardy entropy in the HL quantum gravity model, implying in turn that the corresponding gravitational fields can be analyzed by using the holographic approach of the Cardy entropy. Moreover, the fact that a particular entropy can be associated to homogeneous spacetimes indicate that a thermodynamic analysis can also be performed. This, however, was not analyzed in the present work.

Our results show that the Cardy entropy for homogeneous HL cosmologies depends explicitly on the Ho\v rava parameter $\lambda$. We use this byproduct of our analysis to investigate the physical significance of this parameter. In the case of FLRW spacetimes, we obtained that if $\lambda$ takes values within the interval $[1/3,\infty)$, the Cardy entropy is a real quantity. For $\lambda<1/3$, the entropy becomes imaginary, corresponding to a theory in which only repulsive gravity is present.
We interpret this limiting theory as unphysical. The point $\lambda=1/3$ corresponds to the transition between repulsive and attractive gravity. At this particular point, the Cardy entropy vanishes. Then, for increasing values of $\lambda$ the entropy increases as $S_C\propto \sqrt{\lambda}$. The particular value $\lambda =1$ corresponds to the limit of general relativity at which covariance is recovered. Thus, we can interpret $\lambda$ as the parameter responsible for the intensity of the gravitational interaction in different gravity models.  As $\lambda$ increases and tends to infinity, the HL quantum gravity model describes situations in which quantum effects become more important.

We also calculated the explicit expression for the Cardy entropy of anisotropic spacetimes. The entropy depends explicitly
on $\lambda$ also as $S_C \propto \sqrt{\lambda}$. We investigated in detail the results for the Bianchi IX model which is characterized by three different scale factors. Within the interval $\lambda\in [1,\infty)$, the Hubble parameters can be arbitrarily large, indicating that the velocity expansion in the spatial directions is not limited by the reality condition of the Cardy entropy.  However, in the interval $\lambda\in (1/3,1)$, the situation is completely different. If we fix two of the Hubble parameters of the Bianchi IX model, the remaining third Hubble parameter cannot take an arbitrary value, but is limited within the interval $[0,H_{max}]$
where $H_{max}$ depends explicitly on the value of $\lambda$. This means that an anisotropic spacetime cannot expand with arbitrary velocity along the spatial dimensions. The parameter $\lambda$ acts as a physical barrier that limits the expansion velocity.

We also proved that the Hamiltonian constraints of homogeneous HL spacetimes can be written in a Pythagorean form, leading to a explicit expressions for the Hubble, Bekenstein and Bekenstein-Hawking entropies.  This particular form of the Hamiltonian constraint allows us to  represent the dynamical evolution of the entropies as being inside a circle of radius $S_B$. In contrast to the case of the FLRW spacetime in general relativity, where the radius of the circle is constant, in the anisotropic case   it depends
explicitly on time so that the evolution of the remaining entropies is no longer restricted to be inside a fixed circle.
The radius $S_B$  also depends on the parameter
$\lambda$. This means that for a given value of $\lambda$ the radius of the circle cannot be arbitrarily small or arbitrarily large as time passes, but it
is limited by the value of $\lambda$. We conclude that the Ho\v rava parameter $\lambda$ drastically changes the evolution of the entropies in the Verlinde representation.

Our results show that the holographic approach as expressed in the Cardy entropy can be used in the framework of the HL quantum gravity model to investigate the thermodynamic properties of homogeneous spacetimes. Here, we have only investigated the resulting expressions for the Cardy entropy in order to analyze the physical significance of the Ho\v rava parameter $\lambda$. A more detailed thermodynamic analysis is beyond the scope or the present work. Nevertheless, we consider that our results are a further indication to the possible existence of a not-yet discovered deep relationship between gravity, thermodynamics and holography.

\section*{Aknowledgements}

O.L. wants to thank the National Research Foundation (NRF) for financial support and prof. P.K.S. Dunsby for fruitful discussions. This work was partially supported by DGAPA-UNAM, Grant No. 113514, and Conacyt, Grant No. 166391.

\appendix*

\section{Dynamical equations for  the Bianchi spacetimes}

The three dynamical equations for each Bianchi model can be obtained directly from the HL field equations, and can be written as follows
\begin{equation}
\label{dyneq}
(\lambda-1)\left[-\frac{1}{2}\left(H^{2}_{i}+H^{2}_{j}+H^{2}_{l}\right)+\frac{\ddot{a}_{i}}{a_{i}}+H_{i}H_{j}+H_{i}H_{l}\right]+\lambda\left[H_{j}H_{l}+\frac{\ddot{a}_{j}}{a_{j}}+\frac{\ddot{a}_{l}}{a_{l}}\right]+\frac{k+D}{a_{i}^{2}}=8\pi Gp_{tot}\,,
\end{equation}
where the subindices $\left\{i,j,l\right\}$ take the value $\left\{1,2,3\right\}$, respectively, and the remaining two equations can be obtained by permutation.
The function $D=D(a_{i}, a_{j}, a_{l}, \lambda, \mu)$ is different for each Bianchi model and its explicit value is given in Table \ref{table5}.

\begin{table}
\begin{center}
\begin{tabular}{|c|c|c|}
\hline
Bianchi type & $\qquad k \qquad$ & $D$
\tabularnewline
\hline
& &
\tabularnewline
I & $0$ & $0$
\tabularnewline
& &
\tabularnewline
V & $-1$ & $-\frac{2}{3\lambda-1}\left(\frac{4\pi G\mu}{a_{i}}\right)^{2}$
\tabularnewline
& &
\tabularnewline
&  & $\qquad \epsilon_{i}^{2}-2\xi_{i}^{2}-\frac{2}{3\lambda-1}\left(\frac{4\pi G\mu}{a_{i}}\right)^{2}\left\{4(1-3\lambda)\left[-9\left(1+\frac{\epsilon^{2}_{i}}{3}\right)^{2}+\right.\right. \qquad$
\tabularnewline
IX & $+1$ &
$\left.\left.+6\left(1+\frac{\theta^{4}_{i}}{3}\right)+4\left(1+\xi^{2}_{i}\right)^{2}\right]-\right.$
\tabularnewline
 &  & $\left.-3(1-4\lambda)\left[4\left(1+\frac{\epsilon_{i}^{2}}{3}\right)\left(1+\xi_{i}^{2}\right)-3\left(1+\frac{\epsilon_{i}^{2}}{3}\right)^{2}\right]\right\}$
\tabularnewline
&  &
\tabularnewline
\hline
\end{tabular}
\caption{Dynamical Bianchi equations}
\label{table5}
\end{center}
\end{table}
where
\begin{equation}
\label{xidef}
\xi_{i}^{2} \equiv 1-\frac{a_{l}^{2}}{a_{j}^{2}}-\frac{a_{j}^{2}}{a_{l}^{2}}\left(1-\frac{a_{i}^{4}}{a_{j}^{4}}\right)\, , \qquad \qquad \qquad \,\,\,\,
\end{equation}
\begin{equation}
\label{epsilondef}
\epsilon_{i}^{2}\equiv1-\frac{a_{l}^{2}}{a_{j}^{2}}-2\left(1-\frac{a_{i}^{2}}{a_{j}^{2}}\right)-\frac{a_{j}^{2}}{a_{l}^{2}}\left(1-\frac{a_{i}^{2}}{a_{j}^{2}}\right)^{2},
\end{equation}
\begin{equation}
\label{thetadef}
\theta_{i}^{4}\equiv1-\frac{a_{l}^{4}}{a_{j}^{4}}-2\left(1-\frac{a_{i}^{4}}{a_{j}^{4}}\right)-\frac{a_{j}^{4}}{a_{l}^{4}}\left(1-\frac{a_{i}^{4}}{a_{j}^{4}}\right)^{2}.
\end{equation}


\begin{thebibliography}{99}

\bibitem{prima}
S. Carlip, Rept. Prog. Phys. {\bf 64}, 885 (2001); C. Kiefer, Annal. Phys.  {\bf 15}, 129 (2005).

\bibitem{secondo}
D. Colladay, V.A. Kostelecky, Phys. Rev. D {\bf 58}, 116002 (1998).

\bibitem{terzo} 
G. Amelino-Camelia, Liv. Rev. Rel.  {\bf 16}, 5 (2013); C. J. Isham, \emph{Canonical Quantum Gravity and the Problem of Time}, ArXiv[gr-qc]:9201011 (1992); 
T. Kifune, Astroph. J. Lett. {\bf 518}, 21 (1999); G. Amelino-Camelia, T. Piran, Phys. Lett. B, {\bf 497}, 265 (2001); J. Ellis, N. E. Mavromatos, D. V. Nanopoulos, Phys. Rev. D, {\bf 63}, 124025 (2001).

\bibitem{quarto} .
D. Mattingly, Liv. Rev. Rel., {\bf 8}, 5 (2005); G. Amelino-Camelia, {\it et al.}, Nature, {\bf 393}, 763 (1998); G. Amelino-Camelia, T. Piran, Phys. Rev. D, {\bf 64}, 036005 (2001); G. Amelino-Camelia, Phys. Lett. B, {\bf 528}, 181, (2002).

\bibitem{quinto} 
F. Cianfrani, O. M. Lecian, G. Montani, \emph{Fundamentals and recent developments in nonperturbative
canonical Quantum Gravity}, ArXiv[gr-qc]:0805.2503; S. Coleman, S. L. Glashow, Phys. Rev. D, {\bf 59}, 116008, (1999); J. Collins, {\it et al.}, Phys. Rev. Lett., {\bf 93}, 191301, (2004).

\bibitem{sesto}
U. Jacob, T. Piran, JCAP, {\bf 031}, 0801, (2008); M. Rodriguez Martinez, T. Piran, JCAP, {\bf 006}, 0604, (2006); J. R. Ellis, N. E. Mavromatos, D. V. Nanopoulos, A. S. Sakharov, E. K. G. Sarkisyan, Astropart. Phys., {\bf 25}, 402, (2006).


\bibitem{horava}
P. Ho$\check{\textrm{r}}$ava,  Phys. Rev. D, {\bf 79}, 084008, (2009), P. Ho$\check{\textrm{r}}$ava,  Phys. Rev. Lett., {\bf 102}, 161301, (2009).

\bibitem{teoriadilif}
E. M. Lifschitz, Zh. Eksp. Teor. Fiz., {\bf 11}, 255, (1941).

\bibitem{fq07}
F. J. Hernandez, H. Quevedo, Gen. Rel. Grav. {\bf 39}, 1297, (2007).

\bibitem{ver}
E. Verlinde, \emph{On the holographic principle in a radiation dominated universe}, ArXiv[hep-th]:0008140, (2000).



\bibitem{BlackHoles}
M. I. Park, JHEP, {\bf 123}, 0909, (2009).

\bibitem{GeneralizzazioneHorawa}
H. Lu, J. Mei, C. N. Pope, Phys. Rev. D, {\bf 89}, 084008, (2009); H. Nastase, \emph{On IR solutions in Horava gravity theories}, ArXiv[hep-th]: 0904.3604, (2009).

\bibitem{adddreee}
R. G. Cai, N. Ohta, Phys. Rev. D, {\bf 81}, 084061, (2010).

\bibitem{Muinpark}
M. I. Park, JCAP, {\bf 001}, 1001, (2010).



\bibitem{opq03}
O. Obreg\'on, L. Pati\~no, H. Quevedo, Phys. Rev. D, {\bf 68}, 026002, (2003).


\bibitem{qs}
H. Quevedo and R. Sussman, Jour. Math. Phys., {\bf 36}, 1365, (1995).

\bibitem{kqs}
A. Krasinsky, H. Quevedo, R. Sussman, Jour. Math. Phys., {\bf 38}, 2602, (1997).

\bibitem{suss}
L. Susskind, J. Math. Phys., {\bf 36}, 6377, (1995).

\bibitem{Cardy}
J. L. Cardy, Nucl. Phys. B, {\bf 270}, 186, (1986).

\bibitem{youm99}
D. Youm, Phys. Rep., {\bf 316}, 1, (1999).

\bibitem{sotrep}
T. P. Sotiriou, J. Phys. Conf. Ser., {\bf 283}, 012034, (2011).






\end{thebibliography}
\end{document}